\documentstyle[12pt]{article}
\begin{document}
\date{}
\title{Phase-diagram for Irregular and  Non-Symmetric Cross-linked
 Polymer Blends}
\author{Edilson Vargas and Marcia C. Barbosa \\
Instituto de F\'{\i}sica, Universidade Federal
do Rio Grande do Sul\\ Caixa Postal 15051, CEP 91501-970
, Porto Alegre, RS, Brazil}
\maketitle
\begin{abstract}
We consider here a blend made of two  types of polymers,
$A$ and $B$, of different chemical nature. At high temperature
the homogeneous mixture is cross-linked. As the temperature 
is lowered, the two species try to segregate but are 
kept together by the cross-links. We show that
for 
inhomogeneous, non-regular 
and non-permanent cross-links,  there is  a complete segregation
at low temperatures 
if the system is just weakly cross-linked and  
partial segregation, otherwise. 
We also demonstrate that there is no phase transition between
the homogeneous phase and the microphase
for non-symmetric systems. Our analysis is checked with
the experiment.     
\end{abstract}

\section{\bf Introduction}

The phase separation of polymer blends is
an interesting problem for pratical \cite{Br88}-\cite{Br93}
and fundamental reasons \cite{De77}-\cite{Be97}.
Polymers are present in many materials and the understanding 
of their behavior under changes in temperature, pressure or
magnetic field becomes quite important. 
An interesting case is the mixture of two types of polymer, $A$ and 
$B$, forming a gel. Usually,   chemically different species
are not compatible in the molten state and, consequently, 
 at low temperature this system  segregates 
in two regions, one rich in $A$ and another rich in $B$.
This actually happens for some polymer mixtures
 where the chemical groups
forming one polymer  do not react with the
 compounds forming the monomers of the other.
However,  for most  mixtures this is not the case.
If the two types of polymers are brought into contact
at high temperatures where the entropic free energy 
guarantee their coexistence, the monomers of the polymer $A$ 
react with the monomers of $B$
forming cross-links  \cite{Br88}\cite{Br93}.
When this system is cooled, a 
competition between the natural tendency
for  phase separation and the elasticity of the network
that resists to this separation is established. As a result of 
this competition there is a formation of microdomains alternatively
rich in $A$ and in $B$, what is called microphase
\cite{De79}. The case of a regular and strongly
cross-linked mixture of two species of polymer, $A$ and $B$,
where the polymers where considered to be symmetric and with
the same degree of polimerization, $N$ ( the number
of monomers in each chain), was studied by de Gennes \cite{De79}.
He found that the size of each domain is given by
$\xi\propto a\sqrt{n}$ where $n$ is the 
number of monomers
between two successive cross-links and where
$a$ is the size of each monomer. In his approach, the
position of each cross-link 
does not fluctuate in space. Besides, since they
were homogeneous distributed, $n$ is fixed for the whole
gel. From his analysis, one obtains that the
transition  between the homogeneous phase
and the microphase occurs at a critical temperature
given by $T_c=T_0/(1+\sqrt{6} N/n)$ that is lower than
the temperature $T_0$ where complete segregation would had taken
place\cite{De77}\cite{De79}. For weakly cross-linked mixtures, the 
trapped entanglements 
have to be taken into account \cite {Be91}.
When the number of monomers between two successive
cross-links becomes greater than the 
number of monomers between two successive
entanglements, $n_e$,  the reptation is
the mechanism that leads to the critical  
behavior and, consequently, the size of the microphase is given by
$\xi\propto a\sqrt{n_e}$ \cite{Be91}.  

This is  the scenario for the  phase behavior of these mixtures when
the cross-links are  
{\it  permanent}, {\it regularly distributed} and {\it fixed} in space. In 
 this paper, we will consider a mixture of  
{\it two chemically different} polymers, $A$ and $B$, {\it not compatible}
in the molten state and forming {\it non permanent}, 
{\it not regularly distributed} and {\it mobile}
cross-links. The chemical differences between monomers
$A$ and $B$ means that
regions rich in $A$ and regions rich in $B$ are not 
symmetric and that the two species
exhibit different chemical potentials. In usual critical systems, 
this  does not 
change significantly the phase-diagram, however, in this  particular case,
it leads to the formation of two distinct microphases
and to the absence of phase transition between the
homogeneous and the microphases. Another
effect that we are also taking into account is that
the cross-links are not permanent. Physically this
is relevant for interpenetrating gels where 
the chemical bonds  open and 
reconnect again. In this sense, the cross-links do not differ from the
entanglements that control the physics of non cross-linked mixtures.
Indeed the mobility of the cross-links 
weaken the elasticity of the network increasing
the overall tendency for phase-separation. 
We will also relax the constraint of regularity. 
An uniform distribution of cross-links
 is a good approximation just for
very strong gels, where almost all 
the monomers are cross-linked \cite{De79}\cite{Be91}.
In real systems,
the cross-links are not uniformly distributed along the gel
but according a distribution that differs from system
to system. Here we will represent these inhomogeneities 
in the cross-links  using a Poisson distribution.
In the next section,   we obtain the phase-diagram associated with
our model  that is a  generalization
of de Gennes's approach.

\section{\bf Phase-Diagram  }

In order to investigate the melt of a non compatible 
and {\it non cross-linked} mixture
of  polymers $A$ and $B$ one can  use the 
Landau-Ginsburg-Wilson-de Gennes Hamiltonian 
that contains the entropic free energy associated
with the mixing of the two species and a term related to
the repulsion between the two types of polymer
 \cite{De77}\cite{De79}\cite{Sc94}-\cite{Va97}
\begin{eqnarray}
\label{H0}
\beta H_{0}&=&\int_{}^{} d^{3}r 
\{ a^2\frac{(\nabla \phi(\vec{r}))^2}{48}+\frac{t}{2}\phi(\vec{r})^2+
u\phi(\vec{r})^4-h(r)\phi(\vec{r})\}
\end{eqnarray}
where $a$ is the size of one monomer. In the network, 
the average
concentrations of the two species, 
 $\langle \phi_A(\vec{r}) \rangle=0.5+\langle l(\vec{r}) \rangle$
and  $\langle \phi_B(\vec{r}) \rangle=0.5-\langle l(\vec{r}) \rangle$
are fixed but can have local fluctuations associated with $\phi(\vec{r})$
by
\begin{eqnarray}
\label{phi}
\phi_A(\vec{r})&=&\frac{1}{2}(1+l(\vec{r})+\phi(\vec{r})), \;\; and \\
\phi_B(\vec{r})&=&\frac{1}{2}(1-l(\vec{r})-\phi(\vec{r}))
\end{eqnarray}
For a symmetric medium, $l(\vec{r})=0$, the
difference between the chemical potentials of the two species
also vanishes and, since the linear term in Eq.~(\ref{H0})
does depend on $l(\vec{r})$ and on the chemical potentials, $h(\vec{r})=0$. 
In this case, the 
phase-diagram exhibits a high temperature
disordered phase where $\phi(\vec{r})=0$ what means that the two species
are mixed and a low temperature ordered phase, where $\phi(\vec{r})\neq 0$
what means that the two species are separated. We are going to 
consider  a non-symmetric  medium for which
$\langle l(\vec{r}) \rangle\neq 0$  and where the two chemical
potentials differ and so  $h(\vec{r})\neq 0$.

Besides the chemical differences between the two species
we are going to allow for the $A$ and $B$ polymers to make cross-links
And so, in order to account for  the
elastic forces due to the cross-links, besides the terms in Eq.~(\ref{H0}),
one has to introduce one new term in Eq.~(\ref{H0}).
For finding this contribution, we will use the electrostatic
description proposed by de Gennes \cite{De79} that goes as follows.
 In a dielectric the negative and positive charges 
are tied together but can be displaced. This leads to the appearance of
a polarization. Similarly, in polymer mixtures,
the monomers $A$ and $B$ are linked but when they 
are not fixed, a small displacement of 
their center of masses leads to an elastic
``polarization'' given by
\begin{eqnarray}
\label{P}
\vec{P}&=&\frac{1}{V}(\sum_{i\in A}^{}\vec{r}_i - \sum_{j\in j}^{}\vec{r}_j)
\end{eqnarray}
 where $\vec{r}_i$ is the position of the $i$ monomer at a polymer
 of type $A$  while $\vec{r}_j$ is the position of the $j$ monomer
 of type $B$ and where $V$ is the total volume of the system.
 In the same way
 that for  an electrostatic case, polarization and charge are not
 independent quantities
 , here the elasticity and the volume fraction of each specie
 are also related by
\begin{eqnarray}
\label{nablaP}
\nabla \cdot \vec{P}&=&\phi(\vec{r})+l(\vec{r}).
\end{eqnarray}

Now, using the above description, one has to add to the Hamiltonian Eq.~(\ref{H0})
an elastic contribution  associated with 
the cross-links. For simplicity, we assume that
this term has a quadratic form
that resembles the energy of a spring system namely 
\begin{eqnarray}
\label{Hp}
\beta  H_p&=&\int_{}^{} d^3 r \frac{C(\vec{r}) P(\vec{r})^2}{2}.
\end{eqnarray}
where here $C(\vec{r})$ is the internal rigidity. Within de Gennes approach,
this quantity is a constant and given by 
$C= 36/(n a^2)$.
In our model, however, the
 cross-links are not homogeneous. Inhomogeneities in the cross-links
can be taken into account by assuming that the elastic constant
is a function of the position \cite{Sc94}\cite{St94}
given by
\begin{eqnarray}
\label{C}
C(\vec{r})&=&C_0\sum_{\vec{r}_i}^{}\delta(\vec{r}-\vec{r}_i)
\end{eqnarray}
where the vectors $\{\vec{r}_i\}$ correspond 
to coordinates of $N_c$ particles randomly 
distributed in the volume $V$. These $N_c$ particles are actually the
number of cross-links that are so far distributed
 according a Poisson distribution
characterized by
\begin{eqnarray}
\label{CC}
\langle C(\vec{r}_1)C(\vec{r}_2) \rangle = C_{0} \langle C(\vec{r}_{1}) 
\rangle\delta(\vec{r}_1-\vec{r}_2)
\end{eqnarray}
Since the cross-links can open and close, the disorder is
assumed to be annealed. Consequently the resulting effective hamiltonian
is given by
\begin{eqnarray}
\label{H}
\beta H_{eff}&=&\beta H_{0}+g^{-1}\int_{}^{} d^3 r [1- e^{C_{0} P(r)^{2}/2}].
\end{eqnarray}
where $g^{-1}=N_c/V$. Then, one has to use the
constraint Eq.~(\ref{nablaP}) in order to eliminate
the $\vec{P}$ from the above hamiltonian.
Since $l(\vec{r})$ is 
assumed to be small, its contribution in Eq.~(\ref{nablaP})
just leads to a shift in the linear term $h_q$ in the
expression for $H_{eff}$
and then, the thermodynamic behavior of
the system is  all contained in
the Helmoltz free energy $\beta F_{eff}=- \ln Z_{eff}$
where $Z_{eff}$ is the partition function 
associated with the effective Hamiltonian
$H_{eff}$. The expression for $F_{eff}$ can be evaluated at the
mean-field level by taking the saddle point approximation of
the integral related to  the partition function. 
This approximation
leads  to an effective free energy given by
\begin{eqnarray}
\label{F}
\beta F_{eff}&=&\frac{1}{2}[t+\frac{(q_ca)^2}{24}]\psi_{q_c}\psi_{-q_c}
+u\psi_{q_c}^2\psi_{-q_c}^2
-h_{-q_c}\psi_{q_c}+g[1-e^{- cg\psi_{q_c}\psi_{-q_c}/(2q_c^2)}]\nonumber
\\
\end{eqnarray}
where we have Fourier transformed $Z_{eff}$.
Here
the expressions for 
 $\psi_{q_c}$ and $q_c$ are given by the saddle point equations
\begin{eqnarray}
\label{dF1}
\frac{\partial \beta H_{eff}}{\partial{\phi(q)}}\mid_{\phi_q=\psi_{q_c},q=q_c}
&=&[t+\frac{(q_c a)^2}{24}]\psi_{-q_c}+4u\psi_{-q_c}^2\psi_{q_c}
-h_{-q_c}\nonumber\\
&+&\frac{c}{q_c^2}\psi_{-q_c}e^{- cg\psi_{q_c}\psi_{-q_c}/(2q_c^2)}
\end{eqnarray}
and
\begin{eqnarray}
\label{dF2}
\frac{\partial \beta H_{eff}}{\partial{q}}\mid_{\phi_q=\psi_{q_c},q=q_c}&=&
a^2\frac{q_c }{24}\psi_{q_c}\psi_{-q_c}-\frac{c}{q_c^3}
e^{-cg\psi_{q_c}\psi_{-q_c}/(2q_c^2)}
\end{eqnarray}
and where $c=g^{-1}C_0$.
From the above equations we can see that the system exhibits four possible
phases :

$(a)$ \underline{phase $I$}, a {\it homogeneous phase} 
where $\psi_I\rightarrow 0$ as 
$h_{q_c}\rightarrow 0$ and  where $q_c=q_I\neq 0$;;

$(b)$ \underline{phase $II$},
a {\it complete segregated phase},  where $\psi_{II}\not\rightarrow 0$ as 
$h_{q_c}\rightarrow 0$ and where $q_c=q_{II}=0$;

$(c)$\underline{ phases $III_+$ and $III_-$}, two
microphases where
{\it partial segregation} occurs, where $\psi_{III}\not\rightarrow 0$ as 
$h_{q_c}\rightarrow 0$ and where $q_c=q_{III}\neq 0$.

The free energy associated with each one of these phases is given by
\begin{eqnarray}
\label{FI}
\beta F_I&=&\frac{1}{2}[t+\frac{(a q_I)^2}{24}]
\psi_I^2+u\psi_I^4+g^{-1}[1-e^{-[cg\psi_I^2/(2q_I)]}]
-h_{-q_I} \psi_I
\end{eqnarray}
for the phase $I$,
\begin{eqnarray}
\label{FII}
\beta F_{II}&=&\frac{t}{2}\psi_I^2+\frac{u}{4}\psi_I^4+g^{-1}
\end{eqnarray}
for the phase $II$ and
\begin{eqnarray}
\label{FIII}
\beta F_{III}&=&\frac{1}{2}[t+\frac{(aq_{III})^2}{24}]\psi_{III}^2
+u\psi_{III}^4+g^{-1}[1-e^{-[cg\psi_{III}^2/(2q_{III})]}]-h_{-q_{III}} \psi_{III}
\nonumber \\ 
\end{eqnarray}
for phases $III{\pm}$, Here
the values of $\psi_I,\psi_{II},\psi_{III},q_{I}$ and $q_{III}$
are given by the saddle point solutions of Eq.~(\ref{dF1}) and 
Eq.~(\ref{dF2}).

Then, by comparing the free energies associated
with each one of the phases, we find the  phase-diagram illustrated at
the figure $1$ \cite{Va97} that goes as follows. At high temperatures, 
only the 
homogeneous phase is present. For strong gels ( low $g$ ), as 
the temperature is decreased, 
the microphases  predicted by de Gennes appears. However,
differently from his analysis, there is \underline{no} 
transition between 
the phase $I$ and the phases $III_+$ or $III_-$. As
the temperature is decreased even further, the 
system segregates completely. The transition
between phases $III_{\pm}$ and
$II$ is first-order. For
weak gels, the microphase is not present. Indeed, there is a first-order
phase transition between the homogeneous phase, $I$, and the
completely segregated phase, $II$. 
In our phase-diagram, de Gennes model corresponds to the continuous
transition at ($g=0$, $h=0$). At the plane $h=0$ \cite{Be94}\cite{Sc94},
the critical line, $\lambda$ meets the first-order phase boundaries, $\sigma_{\pm}$,
at the end point $e$. 

Recently, it was suggested  that for any system that has
an end point, the phase boundaries near this region
should exhibit universal features 
 related to the nonanalytic behavior of the thermodynamic functions
near the critical $\lambda$ line \cite{Ba91}. We also verified that this
prediction is actually confirmed in our model for cross-linked
polymer blends  \cite{Va97}.

\bigskip
\bigskip

\section{\bf Discussions and Conclusions}

\bigskip
\bigskip

We have generalized de Gennes-Schulz's model \cite{De79}\cite{Sc94}
 for microphase separation in cross-linked
polymer mixtures  by taking
into account the asymmetry between the two species of polymer, $A$ and $B$, and 
by considering a non-homogeneous distribution of cross-links that are also
not fixed in space.

Our main results are summarized in the phase-diagram illustrated
in  figure. $1$.
We found that   a
mixture of two  chemically different polymers  at high
temperatures  is in an homogeneous phase. If the
system is strongly cross-linked ( high values of $g$), as the temperature is decreased,
there is the formation  of two possible partial segregated phases
or microphases. These phases are characterized by 
forming small domains rich in
in one type of polymer,  $A$, followed by domains  rich in 
the other specie, $B$ or vice-versa. The symmetry between the two microphases
is broken by the  a difference in chemical potential, for example, or
any " field " that would prefer one phase over the other.
In this asymmetric case, there is no phase transition between the
homogeneous and  one of the microphases. If, however, the temperature
is decreased even further, one finds a first-order
phase transition from partial to complete segregation.
If, on the other hand $g$ is large, the polymers 
are just weakly cross-linked and, as the
temperature is decreased, the system 
segregates in two regions, one rich
in $A$ and another rich in $B$. This transition between
the homogeneous phase and the complete segregated phase
 is first-order. 

 The microphases, phases $III_{-}$ and $III_{+}$,
  are characterized by an average domain size that
 is proportional to $\xi\propto 1/q_{III}$. Close to the 
 critical line, the 
 wavevector $q_{III}$ is related to the elastic constant $c$ by 
 $q_{III}\approx (24c/a^2)^{1/4}$. If the number of 
monomers between two cross-links
 is fixed and given by $n$, one finds
 that the coefficient of internal
rigidity is given by $c=36/(na)^2$ \cite{De79}
and, consequently,  $q_{III}=5.42/(an^{1/2})$. However, if
 we assume that the number of 
 monomers between two cross-links is not fixed but
 given by a distribution  of portion sizes $\{n^{\alpha}\}$,
 then $q_{III}= 5.42 \varrho /a(\langle n\rangle)^{1/2}$ where
 $\langle n \rangle$ is the average distance between two cross-links
 and where $\varrho=(\langle n \rangle \langle 1/n\rangle)^{1/4}$
  is a parameter that depends on the distribution.
Now, one  experimental result gives 
$q_{III}= 2.30/(\langle n\rangle)^{1/2}$ \cite{Br88} what agrees just
qualitatively with de Gennes calculation, but is in a good agreement with
our analysis if the distribution $\{n^{\alpha}\}$ would be
such as  $1/\varrho=2.36$.
 
Another interesting result is that for
 weaker gels the microphase disappears
and the system phase separates what  is also 
observed in real systems.\cite{Br88}\cite{Br93} 

\bigskip
\bigskip
\bigskip

\centerline{\it ACKNOWLEDGMENTS}

\bigskip
\bigskip

This work was supported in part by CNPq - Conselho Nacional de
Desenvolvimento Cient\'{\i}fico e Tecnol\'ogico and FINEP -
Financiadora de Estudos e Projetos, Brazil.

\newpage
\bigskip
\centerline{\bf FIGURE CAPTION}
\bigskip
\noindent Figure$1$ .  Phase-diagram $t \times g \times h$ for a $A$-$B$ polymer bend. The phase
$I$ is the homogeneous phase, phase $II$ is the completely segregated phase
and phases $III_{\pm}$ are 
the microphases. The  line $\lambda$, dashed line, is 
a continuous transition,  the planes $\rho$ 
and $\eta$ are  first-order phase boundaries and $e$ locates the end point.
The first-order lines $\sigma_{\pm}$ are
the intersection of the surface $\rho$ with the plane $h=0$.  

\bigskip
\bigskip
\bigskip
\end{document}